\documentclass[
aps,
prb,
reprint,
showkeys,
twocolumn,
nofootinbib,
superscriptaddress
]{revtex4-2}
\usepackage{natbib}
\usepackage{amssymb}
\usepackage{color}
\usepackage{amsfonts}
\usepackage{textcomp}
\usepackage{amsmath}
\usepackage{upgreek}
\usepackage[normalem]{ulem}
\newcommand{\ped}[1]{\ensuremath{_{\rm #1}}}

\definecolor{magenta}{rgb}{0,0,0}
\definecolor{link}{RGB}{57,106,177}
\usepackage{float}
\usepackage[caption = false]{subfig}
\usepackage{graphicx}
\usepackage{hyperref}
\hypersetup{
	colorlinks=true,
	linkcolor=link,
	citecolor=link,
	urlcolor=link
}
\usepackage{xr}

\begin{document}
	
\title{Nodal multigap superconductivity in the anisotropic iron-based compound RbCa$_2$Fe$_4$As$_4$F$_2$}

\author{Daniele Torsello}\thanks{These authors contributed equally.}
\affiliation{Department of Applied Science and Technology, Politecnico di Torino, I-10129 Torino, Italy}
\affiliation{Istituto Nazionale di Fisica Nucleare, Sezione di Torino, I-10125 Torino, Italy}
\author{Erik Piatti}\thanks{These authors contributed equally.}
\affiliation{Department of Applied Science and Technology, Politecnico di Torino, I-10129 Torino, Italy}
\author{Giovanni A. Ummarino}
\affiliation{Department of Applied Science and Technology, Politecnico di Torino, I-10129 Torino, Italy}
\affiliation{National Research Nuclear University MEPhI (Moscow Engineering Physics Institute), Moskva 115409, Russia}
\author{Xiaolei Yi}
\affiliation{School of Physics, Southeast University, Nanjing 211189, People’s Republic of China}
\author{Xiangzhuo Xing}\thanks{These authors contributed equally.}
\affiliation{School of Physics, Southeast University, Nanjing 211189, People’s Republic of China}
\author{Zhixiang Shi}
\affiliation{School of Physics, Southeast University, Nanjing 211189, People’s Republic of China}
\author{Gianluca Ghigo}
\email{Corresponding author: gianluca.ghigo@polito.it}
\affiliation{Department of Applied Science and Technology, Politecnico di Torino, I-10129 Torino, Italy}
\affiliation{Istituto Nazionale di Fisica Nucleare, Sezione di Torino, I-10125 Torino, Italy}
\author{Dario Daghero}\thanks{These authors contributed equally.}
\affiliation{Department of Applied Science and Technology, Politecnico di Torino, I-10129 Torino, Italy}

\begin{abstract}
	The 12442 compounds are a recently discovered family of iron-based superconductors, that share several features with the cuprates due to their strongly anisotropic structure, but are so far poorly understood. Here, we report on the gap structure and anisotropy of RbCa$_2$(Fe$_{1-x}$Ni$_x$)$_4$As$_4$F$_2$ single crystals, investigated by a combination of directional point-contact Andreev-reflection spectroscopy and coplanar waveguide resonator measurements. Two gaps were identified, with clear signatures of $d$-wave-like nodal structures which persist upon Ni doping, well described by a two-band $d{-}d$ state with symmetry-imposed nodes. A large London penetration depth anisotropy was revealed, weakly dependent on temperature and fully compatible with the $d{-}d$ model.
	
	\bigskip
	\textbf{Cite this article as:} 
	
	Torsello, D., Piatti, E., Ummarino, G.A. \textit{et al.} Nodal multigap superconductivity in the anisotropic iron-based compound RbCa\ped{2}Fe\ped{4}As\ped{4}F\ped{2}. \textit{npj Quantum Mater.} \textbf{7,} 10 (2022).
		
	\smallskip
		
	\textbf{DOI:} 
			
	\href{https://doi.org/10.1038/s41535-021-00419-1}{10.1038/s41535-021-00419-1}
\end{abstract}
	
\maketitle
	
\section*{Introduction}

\begin{figure}[h]
	\includegraphics[keepaspectratio, width=\columnwidth]{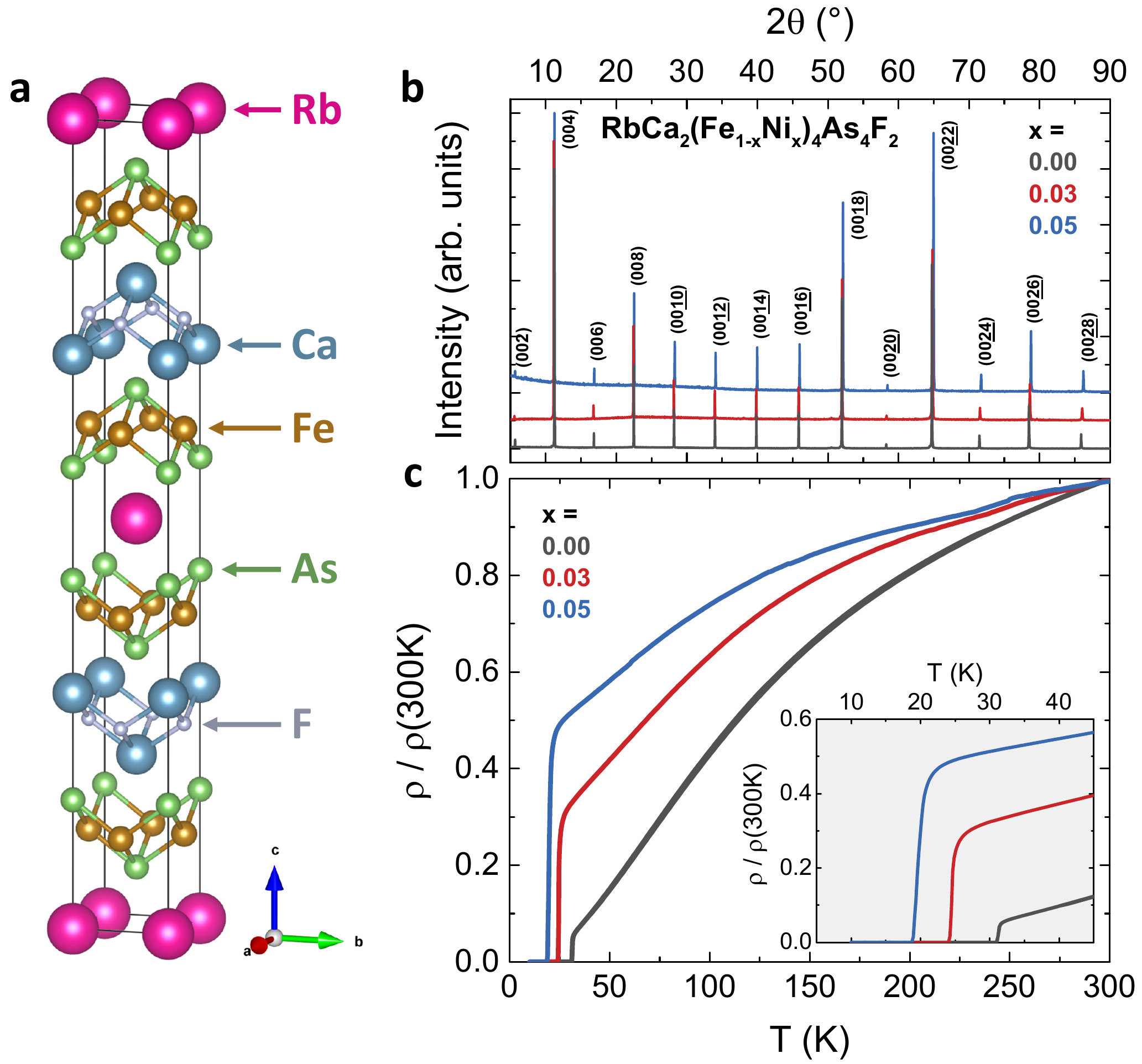}
	\caption{
		\label{Fig_structure}
		\textbf{Structure and basic properties of Rb-12442.} \textbf{a}, Ball-and-stick model of the RbCa$_2$Fe$_4$As$_4$F$_2$ parent compound, as rendered by VESTA~\cite{vesta}.
		\textbf{b}, X-ray diffraction patterns of the differently doped RbCa$_2$(Fe$_{1-x}$Ni$_x$)$_4$As$_4$F$_2$ single crystals studied in this work.
		\textbf{c}, Temperature dependence of the resistivity, normalized by its value at 300~K, of the same single crystals. The inset shows a magnification around the SC transition.}
\end{figure}

The so-called 12442 family of iron-based superconductors (IBSs) is the youngest addition to this class of materials \cite{Wang2016JACS}. It belongs to the generalized 122 structure, in which superconducting (SC) Fe$_2$As$_2$ planes are alternated by spacer layers. In the proper 122 family (such as the AFe$_2$As$_2$ compounds) the spacer layer is made only of A atoms, while in the 1144 systems two 122 layers are alternately stacked. The 12442 family is similar to the 1144 but one of the alternate spacers is not composed of one atomic species but by a larger insulating Ca$_2$F$_2$ layer\,\cite{Wang2017SciChiMat} (Fig.\,\ref{Fig_structure}a). The 12442 structure is therefore less trivial and clearly more anisotropic than those of the rest of the IBSs and, in this respect, more resembling that of cuprates such as Bi$_{2}$Sr$_{2}$CaCu$_{2}$O$_{8+\delta}$ and YBa$_2$Cu$_3$O$_{7-\delta}$ (YBCO) \cite{Yi2020NJP,Wang2019JPCC,Wang2019prb}.
This intuitive and intriguing similarity acquires even more charm in light of the several (although not uncontroversial) experimental observations suggesting the presence of a $d$-wave-like gap opening on the complex, multiband, Fermi surface \cite{Smidman2018PRB,Kirschner2018prb,Wang2020SciChi}. It seems therefore that the 12442 family might be a host of several coexisting features that are key for the understanding of the physics of unconventional superconductors.\\
The Fermi surface of these materials is composed of six to eight sheets (observed by angle-resolved photoemission spectroscopy, ARPES, \cite{Wu2020PRB} or proposed by ab-initio calculations \cite{Ghosh2020}), with large hole pockets at the center of the Brillouin zone and a very small electron pocket at the M point. This situation of almost vanishing electron pockets is close to what is found at the end of the doping series in 122 IBSs (KFe$_2$As$_2$ and KFe$_2$Se$_2$ at the opposite extrema) in which the typical $s_\pm$ symmetry of the order parameter seems not to hold, with the emergence of nodal behavior \cite{Thomale2011PRL,Dong2010PRL,Maiti2012prb,Khodas2012PRL} due to the proximity, in terms of energy, of the two instabilities.\\
Features compatible with the presence of a $d$-wave gap on one of the bands, possibly coexisting with a fully gapped state on other bands, have in fact been observed by $\mu$SR superfluid density measurements (on K- and Cs- 12442 polycrystals \cite{Smidman2018PRB,Kirschner2018prb}) and low temperature specific heat (on K-12442 single crystals \cite{Wang2020SciChi}). Recently, schemes that explain the coexistence of gapless and gapped features were proposed and shown to be suitable candidates for multiband systems \cite{Pang2018,Nica2021}. Conversely, no nodal gaps were observed by heat transport \cite{Huang2019PRB} and optical spectroscopy \cite{Xu2019PRB} in Cs-12442 samples, and by ARPES \cite{Wu2020PRB}, lower critical field measurements\cite{Chu2020CPL} and tunneling spectroscopy (STS) \cite{Duan2021PRB} in K-12442 single crystals. STS measurements revealed a single (contrasting all other reports) $s$-wave gap, with an intraband coupling mechanism that would be incompatible with spin-fluctuation-mediated superconductivity, for which clear signatures were observed by neutron spin resonance measurements \cite{Hong2020PRL,Adroja2020JPCM}. This large set of contrasting reports highlights the complexity of these systems, and the need of joint investigations of the same high-quality samples by means of different techniques in order to shed light on this complicated matter.  \\

In this work, motivated by these considerations, we report on a combined study of the gap symmetry and superfluid density of RbCa$_2$(Fe$_{1-x}$Ni$_x$)$_4$As$_4$F$_2$ single crystals employing directional point-contact Andreev-reflection spectroscopy (PCARS) and a coplanar waveguide resonator (CPWR) technique on samples with increasing Ni doping. The study of doped samples gives the possibility to investigate the persistence of nodal features in the presence of scattering provided by structural disorder, allowing us to speculate and possibly to discern whether they are symmetry-imposed or accidental \cite{Mishra2009prb,Mizukami2014NatCom}.

\section*{Results and Discussion}
\subsection*{Synthesis and basic characterization}
Single crystals of RbCa$_2$(Fe$_{1-x}$Ni$_x$)$_4$As$_4$F$_2$ with $x=0$, 0.03 and 0.05 were grown by the self-flux method using RbAs \cite{Wang2019JPCC,Xing2020sust,Yi2020NJP}. The high quality of the investigated single crystals was assessed through X-ray diffraction, energy dispersive X-ray (EDX) spectroscopy and transport measurements, as shown in \mbox{Fig.\,\ref{Fig_structure}b-c} and in Supplementary Fig. 1. The EDX analysis is performed routinely after crystal growth to determine the exact composition; a typical spectrum and the compositional study are reported in Ref.\,\onlinecite{Yi2020NJP}.

Particularly important is to notice that the resistivity at $T_\mathrm{c}$ increases with increasing Ni content (Fig.\,\ref{Fig_structure}c), indicating an enhanced scattering in the normal state. However, the superconducting transitions remain very sharp despite the strong decrease of critical temperature (inset of Fig.\,\ref{Fig_structure}c). This highlights the primary role of Ni substitution in shifting the chemical potential and changing the electron doping level, rather than enhancing the pair-breaking scattering in the superconducting state (that would broaden the transition). This effect was pointed out also by Hall coefficient measurements in Ref.\,\onlinecite{Yi2020NJP}, although the $T_\mathrm{c}$ dependency on the Ni content was there interpreted as an indirect indication of the $s_\pm$ symmetry of the order parameter. Actually, due to this double role of Ni doping, direct measurements are necessary to investigate fundamental details of the gaps, as we will show in the following.

\begin{figure*}
	\includegraphics[keepaspectratio, width=\textwidth]{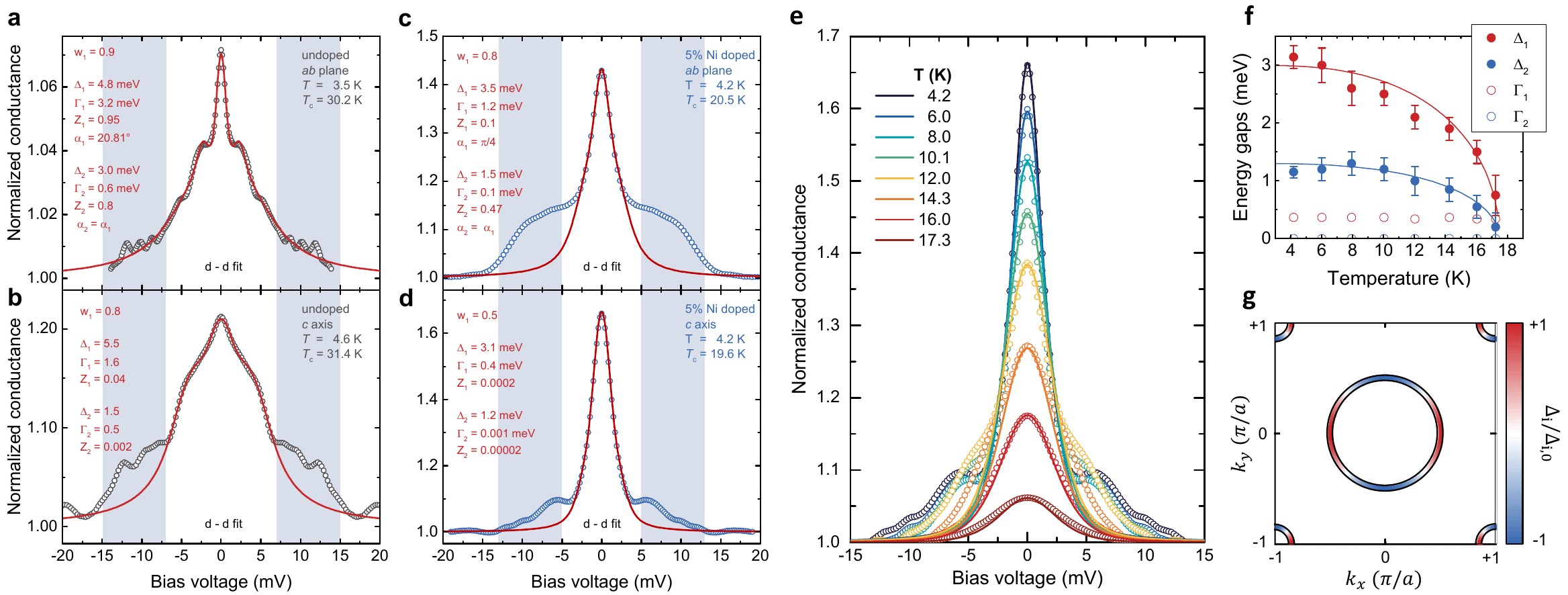}
	\caption{
		\label{Fig_PCARS} 
		\textbf{PCARS spectra and superconducting gap symmetry of Rb-12442.}
		Representative low-$T$ directional PCARS spectra (symbols) on single crystals of RbCa$_2$(Fe$_{1-x}$Ni$_x$)$_4$As$_4$F$_2$ with $x=0$ (\textbf{a},\textbf{b}) and $x=0.05$ (\textbf{c},\textbf{d}), and corresponding two-gap fits of the spectra in a $d{-}d$ model (solid red lines). The fitting parameters are indicated in labels. 
		\textbf{e}, Temperature dependence of the PCARS spectra (symbols) and corresponding $d{-}d$ fits (solid lines) for the same contact shown in \textbf{d}.
		\textbf{f}, Superconducting energy gaps (filled circles) and broadening parameters (hollow circles) as a function of temperature, determined by fitting the spectra shown in \textbf{e} with the $d{-}d$ model. Error bars represent the interval of gap values obtained with different normalizations of the same curve, as explained in the Methods, and include the uncertainty due to the fit. Solid lines are the expected BCS dependencies.
		\textbf{g}, Scheme of the gap structure (normalized gap values as color scale) showing the compatibility of the $d{-}d$ symmetry with the underlying Fermi surface structure.
	}
\end{figure*}

\subsection*{Directional point-contact Andreev-reflection spectroscopy}
The number and symmetry of the SC energy gaps were determined via directional PCARS\,\cite{DagheroSUST2010, Gonnelli2001eEPJB, ZhigadloPRB2018, DagheroPRB2020, DagheroSUST2018} in crystals with ${x=0}$ and ${x=0.05}$. Fig.\,\ref{Fig_PCARS}a-d reports some examples of normalized differential conductance spectra ($dI/dV$, symbols) acquired by injecting the current along either the $ab$ plane or the $c$ axis (see Methods, Supplementary Methods, and Supplementary Fig. 4 for details). As in all the Fe-based compounds, in addition to the structures associated to the energy gap(s), the spectra present higher-energy shoulders (falling in the grey regions) that can instead be associated to the strong coupling between carriers and the bosonic mode responsible for the pairing\,\cite{Tortello2010,Daghero2011,Daghero2014}.\\
In both the undoped and 5\% Ni-doped crystals, 100\% of the spectra display zero-bias peaks or cusps that are strongly suggestive of a gap with vertical node lines and a change of sign on the same Fermi surface \cite{DagheroPuCoGa}. This agrees with all the evidences of nodal gaps in other compounds of the same family (Cs-12442 and K-12442), but is completely incompatible with a single $s$-wave gap picture (as observed in K-12442 by STM\cite{Duan2021PRB}).
The suppression of the critical temperature $T_\mathrm{c}$ by $\sim10$~K in the 5\% Ni-doped samples with respect to the undoped compound results  in a decrease in the gap amplitudes but, also, in a reduction of the energy at which the boson structures set in; as a result, the latter fall very close to the edge of the large gap, which makes it more difficult to disentangle the relevant contribution. \\
To extract quantitative information on the gaps, it is necessary to fit the spectra with a suitable model for Andreev reflection in the presence of a nodal gap. We used the one reported in Ref.\,\onlinecite{KashiwayaPRB1996}, which perfectly fits the case of $ab$-plane contacts, and its generalization to the case of $c$-axis contacts in a highly anisotropic material \cite{YamashiroPRB1997}. Some details about this model and its parameters are given in the Methods and Supplementary Methods. Since it is based on the BCS, weak-coupling assumption of energy-independent order parameters, this model is unable to reproduce the bosonic structures; therefore, the fit has been made by disregarding them and focusing on the central part of the curves, that instead contains all the information about the SC energy gaps; the validity of this approach has been demonstrated in Refs.\,\onlinecite{Daghero2014,Daghero2011}. As shown in Supplementary Fig.~7c-d, if one tries to fit these structures as if they were due to a gap, very high (and probably non-physical) values of the gap ratio $\kappa=2\Delta/k_\mathrm{B}T_\mathrm{c}$ are obtained. 

\subsection*{Gap structure}
As for the gap symmetry, a guess has to be made before trying to fit the spectra. Since specific-heat\,\cite{Wang2013} and $\mu$SR\,\cite{Smidman2018PRB,Kirschner2018prb} measurements in similar compounds were interpreted as suggesting the coexistence of a nodal gap together with a nodeless one, we initially tried to fit the spectra by using one $s$-wave gap and one nodal gap, for simplicity assumed to have a $d$-wave symmetry, reasonably residing on different Fermi surface sheets. We found that all curves for undoped and 5\% Ni doped crystals, both in $ab$-plane and $c$-axis contacts, can be satisfactorily fitted by this $s{-}d$ model, with a \emph{larger} $s-$wave gap $\Delta_1$ and a \emph{smaller} $d-$wave gap $\Delta_2$, but the contribution of the $s$ gap is barely visible and the gap values extracted from different contacts are rather scattered (for $x=0$, $\Delta_1= 4.3 - 6.6$ meV and $\Delta_2= 1.9 - 3.5$ meV) and, within the experimental uncertainty, look to be directly proportional to $T_\mathrm{c}$ (see Supplementary Fig.~8).  

Despite this ostensible success, the existence of a nodal and a nodeless gap in the same system poses some questions from the theoretical point of view, and may not be the most reasonable explanation of the observed results. In particular, doubts might arise regarding the compatibility of this gap structure  with the underlying $I4/mmm$ point group symmetry and Fermi surface structure. Moreover, the electron-electron interaction in the IBSs is mainly provided by repulsive antiferromagnetic spin fluctuations (AFM-SF), that provide coupling over regions of the Fermi surface with opposite sign of the SC gaps. Thus, a single $s$-wave gap cannot exist within this picture, but two isotropic gaps with opposite sign ($s_\pm$ state) are possible. Therefore, in an effective two-band model with $s$- and $d$-wave gaps, one needs to admit that \textit{two} $s$-wave gaps with \textit{opposite sign} open up on either band ($\Delta_{s,1}$ and $\Delta_{s,2}$); as for the $d$-wave gap, it can be considered to exist in both the bands for the sake of generality: $\Delta_{d,1}=\Delta_{d,2}$. In other words, each band should host a $s+d$ gap. This gap displays nodes only if the $d$-wave component is larger than the $s$-wave one, in contrast with the experimental results of the fit (see also Supplementary Fig. 8).  Therefore, a two-band $s{-}d$ model is likely to be, at least, oversimplified for this system. 

Since the PCARS spectra (as well as the CPWR measurements, as discussed below) unambiguously point to the existence of nodes in the gap, it makes sense to fit them by assuming either a single nodal gap (which however does not allow properly fitting all the curves) or \emph{two} nodal gaps. For simplicity, we chose a $d_{x^2-y^2}$-wave symmetry for both of them, with the nodal direction at an angle $\uppi/4$ with respect to the $k_x$ axis in the reciprocal plane (see Methods). Such a two-band $d{-}d$ picture automatically: i) ensures the existence of nodes, ii) respects the overall symmetry of the system, and iii) does not pose problems of coexistence of the different gaps even in the simplest two-band model. In the following, we will show that it is also fully compatible with our PCARS and CPWR results. Fig.\,\ref{Fig_PCARS}g depicts a possible realization of this $d{-}d$ picture in the case where the Fermi surface consists of a holelike sheet at the centre of the Brillouin zone, and of an electronlike pocket at the corner: a schematic simplification of the 12442 case.\\
In general, the quality of the fits of the PCARS spectra within the $d{-}d$ model is comparable to that obtained with the $s{-}d$ model. However, in the cases where the zero-bias peak is the more relevant feature, the $d{-}d$ fit is by far the most natural choice and reproduces very easily the experimental data. In the case of $ab$-plane contacts, for example, the change of sign in the gap is expected to give rise to a constructive interference of electronlike and holelike quasiparticles (ELQ and HLQ in the following) whenever the current is injected at an angle $\alpha\neq 0$ with respect to the $k_x$ axis, with a maximum probability for $\alpha=\uppi/4$. This phenomenon explains both the peak-shoulders feature of the spectrum in Fig.\,\ref{Fig_PCARS}a ($\alpha\simeq \uppi/8$) and the high zero-bias peak of Fig.\,\ref{Fig_PCARS}c ($\alpha=\uppi/4$). In $c$-axis contacts this interference cannot take place since ELQ and HLQ always feel gaps with the same sign, but the existence of nodes makes unpaired quasiparticles to be available for conduction even at very low energy. This naturally explains the zero-bias maxima in Figs. \ref{Fig_PCARS}b,d; the latter looks narrower mainly because of the smaller amplitude of the larger gap.

As a further check of the correctness of our approach, we report in Fig. \ref{Fig_PCARS}e the temperature dependence of the spectrum shown in Fig. \ref{Fig_PCARS}d . The experimental conductance curves (symbols) are compared to the relevant fitting curves (solid lines). We chose to keep all the parameters fixed to their low-temperature values, and to tune only the gap amplitudes. Despite this very strict constraint, the theoretical curve follow very well the evolution of the central part of the spectra. Moreover, the gap amplitudes approximately follow a BCS trend, as shown in Fig. \ref{Fig_PCARS}f.

When the whole series of measurements is considered, it turns out that the gap amplitudes extracted from the $d-d$ fit are less scattered than in the $s{-}d$ case: for $x=0$, $\Delta_1$ ranges between 4.9 and 6.6 meV and $\Delta_2$ from 1.5 to 2.5 meV. 
Fig.\,\ref{Fig_discussion}a reports the gap amplitudes as a function of the $T_\mathrm{c}$ of the contacts; the shaded regions are delimited by lines of constants gap ratio. For the larger gap, $2\Delta_2/k_\mathrm{B}T_\mathrm{c}$ ranges from 3.3 and 5.0 and is centered at about 4.2, the value predicted by the BCS theory in the $d$-wave case. The smaller gap has instead a gap ratio between 1 and 2, much smaller than the BCS value.
Interestingly, the $c$-axis and $ab$-plane gap amplitudes follow the same trend, with no evidence of a dependence on the direction of current injection. 

\begin{figure}
	\includegraphics[keepaspectratio, width=\columnwidth]{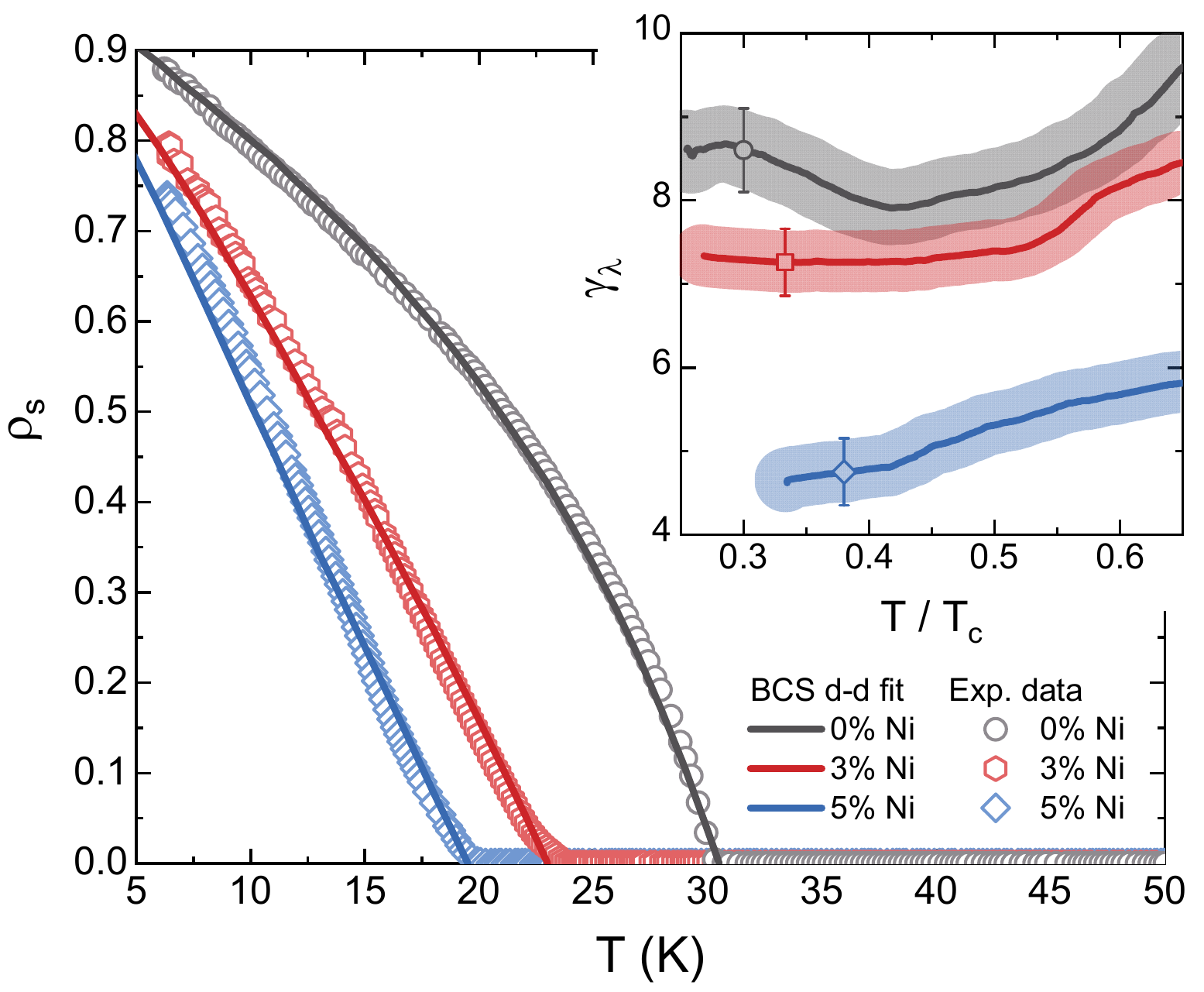}
	\caption{
		\label{Fig_CPWR} \textbf{Superfluid density and anisotropy of Rb-12442.}
		Temperature dependence of the normalized superfluid density measured with the CPWR technique on single crystals with $x=0$, 0.03 and 0.05 (symbols) and their $d{-}d$ BCS fits (lines). 
		Inset shows the London penetration depth anisotropy ($\gamma_\lambda$) as a function of the normalized temperature for the three doping levels. The shaded bands around the lines represent the error of the reported $\gamma_\lambda$ values (an illustrative error bar is also shown, for a single point of each curve), arising from the uncertainties of each $\lambda$ measurement and their propagation.  
	}
\end{figure}

\subsection*{Coplanar waveguide resonator analysis}
\begin{figure}
	\includegraphics[keepaspectratio, width=\columnwidth]{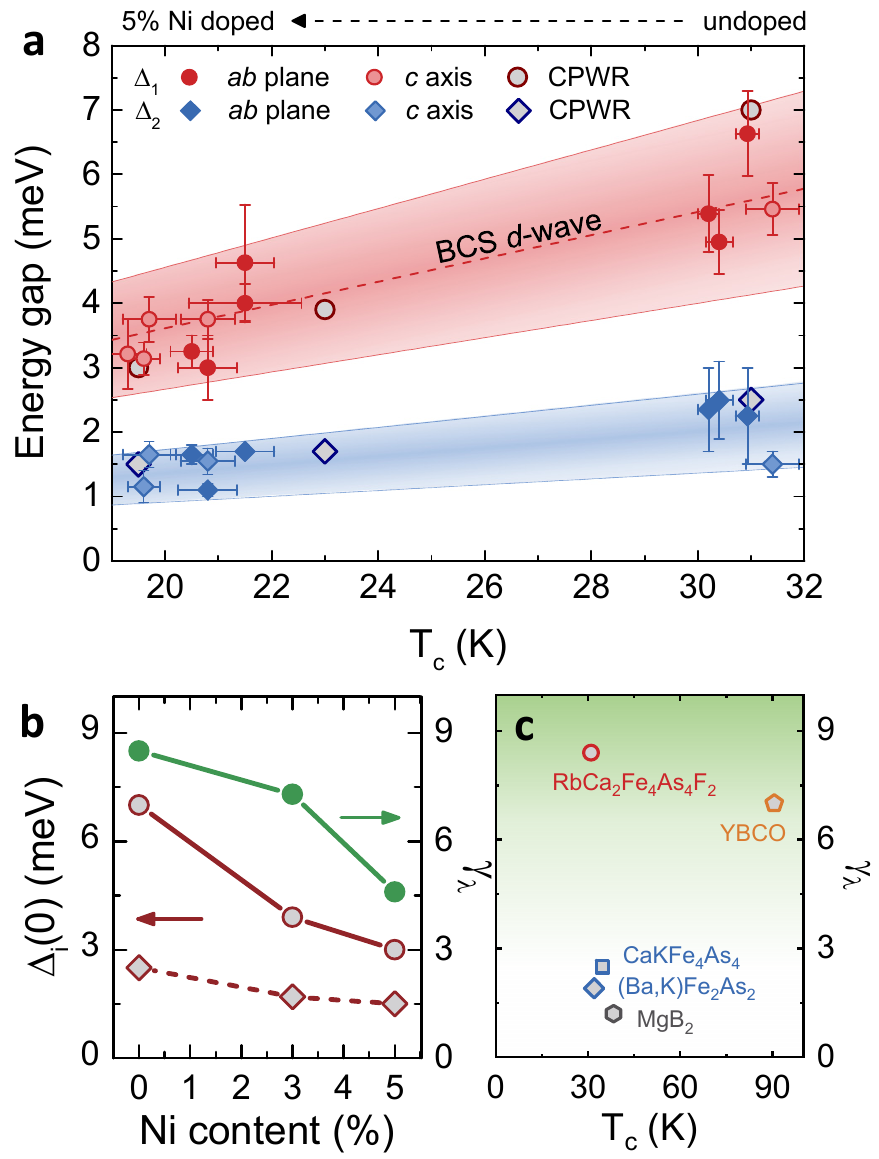}
	\caption{
		\label{Fig_discussion} \textbf{Gap values and anisotropy of Rb-12442.}
		\textbf{a}, Amplitude of the SC energy gaps extracted from the fit of PCARS spectra using the $d{-}d$ model as a function of the Andreev $T_\mathrm{c}$, and from the $d{-}d$ wave BCS fit of the superfluid density. Error bars represent the interval of gap values obtained with different normalizations of the same curve, and include the uncertainty arising from the fit. The colored regions are bound by lines with constant gap ratio $2\Delta/k\ped{B}T_\mathrm{c}$. The dashed line is the gap amplitude expected for a BCS, single-gap, $d$-wave superconductor.
		\textbf{b}, Ni-doping dependence of $\gamma_\lambda$ at 6 K (right scale) and of the SC energy gap values extracted from the BCS fits in Fig.~\ref{Fig_CPWR} (left scale).
		\textbf{c}, Anisotropy of the London penetration depth (measured at $T\simeq5-6$ K) in the undoped Rb-12442 single crystal compared to that of MgB$_2$ \cite{Fletcher2005PRL}, other IBSs of the generalized 122 family~\cite{Khasanov2009PRL, Torsello2019PRAppl}, and YBCO~\cite{Pompeo2015JAP}. }
\end{figure}

As already mentioned, a combination of multiple bands and unconventional gap symmetry in 12442 compounds was also suggested on the basis of the low-temperature linear feature in the superfluid density observed by $\mu$SR in K- and Cs-12442 polycrystals \cite{Smidman2018PRB,Kirschner2018prb}. Fig.\,\ref{Fig_CPWR} shows the temperature dependence of the normalized superfluid density $\rho_\mathrm{s}$ in our undoped (${x=0}$) and Ni-doped (${x=0.03}$ and 0.05) single crystals, determined by starting from the CPWR measurement of the London penetration depth $\lambda$, i.e. $\rho\ped{s}=[\lambda({T=0})/\lambda]^2$ (see Supplementary Fig. 3 and Refs.\,\onlinecite{Ghigo2018sust,Torsello2019prb,GhigoPRR,Torsello2019EPJST}). A linear temperature dependence of $\rho_\mathrm{s}$ is, indeed, evidenced also in our samples, in the experimentally accessible temperature range. The absence of kinks in the superfluid density curve is typical of single band superconductors and of multigap systems with mainly interband coupling, as the IBSs \cite{Torsello2019prb}. The temperature dependence of the superconducting gaps measured by PCARS  (Fig. \ref{Fig_PCARS}f) indeed shows that both gaps slowly decrease and close -together- at $T_\mathrm{c}$, therefore a smooth $\rho_\mathrm{s}(T)$ curve is expected. With respect to Refs.\,\onlinecite{Smidman2018PRB,Kirschner2018prb}, our measurements display a much less pronounced curvature at high $T$, and a wider $T$ range where the behavior is quite linear, particularly evident in Ni-doped samples. These differences might be due to differences among the systems (Cs-, K- and Rb- based) or to the single- vs. poly-crystalline type of samples. Notably, our $\rho_\mathrm{s}$ data can be fitted, for all doping levels, with a BCS models involving two $d$-wave gaps, as done in Refs.\,\onlinecite{Smidman2018PRB,Kirschner2018prb} (details in the Supplementary Methods and Supplementary Table 1), resulting in SC gap values (shown in Fig.\ref{Fig_discussion}a-b) that are in very good agreement with those found experimentally by PCARS. This is a further support to the validity of the $d{-}d$ picture for Rb-12442, but is not in contrast with the $\mu$SR experiments on  Cs- and K-12442\,\cite{Smidman2018PRB,Kirschner2018prb}, since in those systems the $s{-}d$ and the $d{-}d$ fits are also very similar, possibly due to the contribution of several bands to the overall density. It should be noted that, although also a single $d$-wave gap fit would give a reasonable result, the agreement with the PCARS data (unambiguously pointing to the presence of at least two gaps) would be lost.
The fact that the same nodal features are observed, both in PCARS and CPWR measurements, up to 5\% Ni doping is a strong indication that Ni doping, up to this level, neither affects the number and the symmetry of the energy gaps, nor lifts the nodal lines, suggesting that they might be symmetry-imposed and not accidental \cite{Mishra2009prb}. In principle, however, this could also mean that the amount of scattering provided by this doping level is just not sufficient to lift the nodes. In this respect, we recall the above discussion, regarding the weak role of Ni atoms as pair-breaking scattering center. This is further evidenced by the fact that there is no clear change in the $\rho_\mathrm{s}(T)$ behavior at low T, that is expected to change from linear to quadratic for a $d$-wave superconductor in the dirty limit. Additional studies, with disorder introduced via ion or electron irradiation \cite{Ghigo2018PRL,Torsello2020sust,Cho2016SciAdv,Torsello2021SciRep} in undoped samples are desirable and ongoing. 

\subsection*{Anisotropy}
Starting from the CPWR measurements of the London penetration depth of crystals with different shape factor, the penetration depth anisotropy parameter $\gamma_\lambda=\lambda_c/\lambda_{ab}$ can be extracted\,\cite{Torsello2019PRB2,Torsello2019PRAppl} (see Methods and Supplementary Methods). The measured values, shown in the inset to Fig.\,\ref{Fig_CPWR}, are much larger than what is generally found for other IBS of the generalized 122 family, but in line with those reported for the upper critical field in the same system\,\cite{Yi2020NJP} and in K-12442 \cite{Wang2020SciChi_pulsed} and close to the typical values found in cuprates\,\cite{Pompeo2015JAP}, as shown in Fig.\,\ref{Fig_discussion}c.
Moreover, $\gamma_\lambda$ is found to decrease with increasing  Ni substitution (Fig.\,\ref{Fig_discussion}b), as expected due to the presence of Ni impurities, that act also as weak isotropic scattering centers \cite{Kogan2002PRB}. However, at 5\% Ni content the anisotropy parameter value is still about twice what is found for undoped CaKFe$_4$As$_4$ \cite{Torsello2019PRAppl}. The temperature dependence of $\gamma_\lambda$ is quite flat, with a slight upturn at higher temperatures. 
For long, a temperature dependence of this quantity was interpreted as a signature of the multiband nature of the system. However, recent calculations \cite{Kogan2019PRB} have shown that in multiband systems with spheroidal Fermi surfaces any temperature dependence (increasing, decreasing or constant) is achievable depending on several microscopic parameters. In short, these calculations indicate that: i) for a single $d$-wave gap with vertical node lines, of the form $\Delta \cos(2\phi)$ ($\phi$ being the azimuthal angle) $\gamma_\lambda$ should be temperature independent, as in the single-band $s$-wave case; ii) for the multiband case, even two $s$-wave gaps can give rise to any temperature dependence of $\gamma_{\lambda}$, depending on the band anisotropies. 
Following the same approach as in Ref. \onlinecite{Kogan2019PRB}, it is possible to generalize the calculations to a two $d$-wave gap system on spheroidal Fermi surface (details in the Supplementary Methods). One finds that the results discussed in Ref.\,\onlinecite{Kogan2019PRB} for the two-band $s$-wave case are strictly valid also for the two $d$-wave gaps case. Therefore, the weak temperature dependence found for $\gamma_\lambda$ is fully compatible with the $d{-}d$ gap structure considered.
It also follows, from the constancy of the gap ratios measured along the $ab$ and $c$ directions for undoped and doped samples (compared to the strong decrease of $\gamma_\lambda$ with doping) that the main factors influencing the value of $\gamma_\lambda$ could be the anisotropy of the Fermi velocity and the 2D (cylindrical as opposed to spheroidal) topology of the Fermi surfaces. The decrease  of $\gamma_\lambda$  with doping could be determined by isotropic scattering provided by the impurity atoms as in the 1144 system \cite{Torsello2019PRAppl}, or by a change of values of the Fermi velocities that seems reasonable when the Fermi level is shifted by doping.

\subsection*{Concluding remarks}
In conclusion, we have brought evidence for the existence of two nodal (possibly $d$-wave) gaps in Rb-12442 by using two techniques, PCARS and CPWR, that allow investigating the symmetry of the order parameter from different perspectives and yielded concording results. The shape of the PCARS spectra, as well as the low-temperature linear behaviour of the superfluid density measured by CPWR, clearly indicate the presence of nodes in the gap that persist upon Ni doping, suggesting that the nodes might be symmetry imposed. Moreover, both the PCARS spectra and the $\rho_\mathrm{s}(T)$ curve can be fitted very well within a $d{-}d$ model (compatible with the Fermiology and symmetry of the system) yielding gap amplitudes in very good agreement. These facts, together with the unusually high values of the London penetration depth anisotropy, seem to tighten the similarity between the 12442 family of IBSs and the copper oxides, which makes the former an ideal platform to investigate the nature of superconductivity in both these classes of unconventional superconductors.
	
\section*{Methods}	
\begin{footnotesize}
\subsection*{PCARS measurements}\label{S_PCARS}
The superconducting gap structure was investigated via point-contact Andreev reflection spectroscopy (PCARS), which entails the measurement of the differential conductance ($dI/dV$) of a point-like contact between the superconductor (S) and a normal metal (N). Here, the contacts were fabricated via the so-called soft technique\,\cite{DagheroSUST2010, ZhigadloPRB2018, DagheroPRB2020, DagheroSUST2018}, in which a thin Au wire is stretched over the crystal, until it touches the surface in a single point, is (optionally) mechanically stabilized by drop-casting a small ($\approx 50\,\mu m$) droplet of Ag glue, and acts as a parallel of nanoscopic N/S contacts. Thanks to the platelet-like shape of the crystals, it was possible to control the direction of (main) current injection, by making the point contacts either on the topmost surface (thus injecting the current along the $c$ axis) or on the side (thus injecting the current along the $ab$ planes). We refer to these contacts as $c$-axis and $ab$-plane contacts both in the following and in the Results and Discussion section.  Actually, to ensure a good directionality of the current injections, $c$-axis contact were made in flat, mirrorlike regions of the top surface, excluding growth terraces; as for $ab$-plane contacts, the very thin, but smooth and flat side surfaces of the crystals allowed us to assume a small, if any, contribution of conduction along the $c$ axis. A picture of contacts along either direction is reported in Supplementary Fig. 4.

Proper PCARS measurement conditions require the contact to be smaller than the electronic mean free path, resulting in ballistic conduction with no Joule dissipation in the contact region, and a contact resistance largely exceeding the resistance of the normal bank\,\cite{DagheroSUST2010,Daghero2011}. In these conditions -- and in the absence of a surface insulating layer -- electronic conduction across the S/N interface is dominated by Andreev reflection\,\cite{BlonderPRB1982}, and a quantitative analysis can be carried out by fitting the experimental $dI/dV$ vs $V$ spectra with suitable models.

Prior to fitting, the experimental $dI/dV$ curves need to be normalized, i.e. divided by the normal-state differential conductance curve\,\cite{DagheroSUST2010, DagheroPRB2020, DagheroSUST2018}. In principle, the $dI/dV$ spectrum recorded at a given $T < T_\mathrm{c}$ should be divided by the normal-state conductance at the \emph{same} temperature, obtained by suppressing superconductivity with, e.g., a magnetic field. However, when the upper critical field is very high (as in most IBS) a common practice is to use the normal-state conductance recorded just above $T_\mathrm{c}$. The small thickness of the crystals (a few micrometers) however leads to a rather large normal-state resistance that, in the pseudo-four-probe configuration used for PCARS, is in series with the contact resistance and gives rise to a downward shift and a horizontal stretching of the conductance curves with respect to those at low temperature\,\cite{PecchioPRB2013,Daghero2014}. To recover the presumable normal-state conductance at low temperature, we rescaled the $dI/dV$ at $T>T_\mathrm{c}$ by subtracting the contribution of this so-called spreading resistance $R_\mathrm{s}$. The value of $R_\mathrm{s}$ for each contact was typically selected as the one that allowed to match the tails of the superconducting and normal $dI/dV$ spectra, but the choice remains somewhat arbitrary. Therefore, multiple reasonable values of the spreading resistance were used to obtain the normalized spectra, and the fitting procedure repeated each times. The resulting values of the fit parameters were then averaged and their maximum differences taken as the uncertainty of the procedure. 

The normalized PCARS spectra were fitted to a two-band model, using either two $d$-wave gaps (as discussed in the Results and Discussion section), or one $s$-wave gap and a $d$-wave gap (summarized in Supplementary Figs. 7,8). The quasi-cylindrical shape of the Fermi surfaces is taken into account while performing the integration over all possible directions of incident electrons. For $c$-axis contacts, we integrated over a cylindrical belt about the basal plane $k_z=0$, as explained in Ref.\,\onlinecite{YamashiroPRB1997}. For $ab$-plane contacts, we integrated in the whole half-plane $k_x>0$ as explained in Ref.\,\onlinecite{KashiwayaPRB1996}. In either case, the model is actually the same and takes into account the interference effects that occur when HLQ and ELQ (injected with different wavevectors) feel order parameters of opposite sign. This actually occurs only in $ab$-plane contacts, when the normal to the N/S interface makes a non-zero angle $\alpha$ with the (antinodal) $k_x$ direction, and can give rise to zero-energy peaks in the spectra (see Supplementary Methods for more details). 
In both cases the model contains as adjustable parameters the gap amplitudes $\Delta_1$ and $\Delta_2$, the broadening parameters $\Gamma_1$ and $\Gamma_2$, the barrier parameters $Z_1$ and $Z_2$, and the relative weight of the large-gap band in the conductance $w_1$. 
In the $d{-}d$ case, $\alpha$ was supposed to be the same for both gaps. Despite the large number of parameters, the values they acquire in the fits is not arbitrary because each parameter has different and specific effects on the shape of the spectrum, as extensively discussed in Refs.\,\onlinecite{DagheroSUST2010, ZhigadloPRB2018, DagheroPRB2020, DagheroSUST2018}. In spectra with a low signal, as in Fig.\ref{Fig_PCARS}a, there is some interplay between $\Gamma$ and $\Delta$ which gives rise to an increased uncertainty on the gap amplitude, here included in the error bars. In all cases, the high-energy shoulders highlighted by the shaded gray bands in Fig.\,\ref{Fig_PCARS}a-d and Supplementary Fig.~7 must be excluded by the fitting procedure, since they are strong-coupling features that are not captured by models based on the weak-coupling BCS theory\,\cite{Tortello2010,Daghero2011,Daghero2014}. Indeed, if these structures are fitted as if they were due to an energy gap, the resulting gap amplitude is unphysically large (see Supplementary Methods for details).

\subsection*{CPWR measurements}\label{S_CPWR}
The superfluid density, surface impedance and London penetration depth anisotropy were measured by means of a coplanar waveguide resonator (CPWR) technique that has proven suitable to study IBS crystals \cite{Ghigo2017scirep,Ghigo2018PRL,GhigoPRR}. The measurement device consists of an YBa$_2$Cu$_3$O$_{7-\delta}$ resonator in the shape of a coplanar waveguide, to which the sample is coupled. Resonance curves are recorded with a vector network analyzer as a function of temperature, making it possible to track resonant frequency shifts and variations of the unloaded quality factor across the superconducting transition of the investigated samples. After a calibration procedure \cite{Ghigo2017prb,GhigoPRR} is performed, this gives access to the absolute value of the penetration depth and its temperature dependence, as well as to the surface impedance temperature dependence \cite{Torsello2019EPJST,Torsello2019JOSC}.\\ Due to the fields distribution at the sample position (vanishing electric field and rf magnetic field parallel to the $ab$-planes of the sample) and to the finite size of the crystal, the measured penetration depth $\lambda$ is a combination of the main anisotropic components $\lambda_{ab}$ and $\lambda_{c}$. This combination depends on the geometry of the sample under consideration, and for this reason both components can be retrieved by measuring samples with significantly different aspect ratios $f_\mathrm{s}=c\cdot(1/a+1/b)$ \cite{Torsello2019PRAppl}.\\
Accordingly, and for samples with $\lambda_{ab}<c$ and $\lambda_{c}<a, b$ (where $2c$, $2a$, $2b$ are respectively the thickness, width and length of the samples), the fraction of penetrated volume can be estimated as $\lambda_{ab}/c+\lambda_{c}/a+\lambda_{c}/b$. Thus, the measured penetration depth can be expressed as $\lambda=\lambda_{ab}+f_\mathrm{s} \cdot \lambda_{c} $.

\section*{Data Availability}
All data needed to evaluate the conclusions in the paper are present in the paper and/or the Supplementary Methods. Additional data related to this paper may be requested from the authors.

\section*{Acknowledgements}
The authors thank P. J. Hirschfeld, E. M. Nica, Q. Si and A. Tibaldi for fruitful discussions. This work was partially supported by the Italian Ministry of Education, University and Research (Project PRIN HIBiSCUS, Grant No. 201785KWLE and Project PRIN Quantum2D, Grant No. 2017Z8TS5B), by the National Key R\&D Program of China (Grant No. 2018YFA0704300) and the Strategic Priority Research Program (B) of the Chinese Academy of Sciences (Grant No. XDB25000000). G.A. Ummarino acknowledges partial support from the MEPhI.

\section*{Author contributions}
The work was conceived by D.T. and G.G., who also conducted CPWR measurements and analyzed the results. E.P. and D.D. conducted PCARS measurements and analyzed the results, X.Y., X.X. and Z.S. provided the crystals and performed basic characterization, while G.A.U. applied the theoretical models. The manuscript was written by D.T., D.D., E.P. and G.G. with comments and input from all authors. D.T., E.P., X.X, and D.D. contributed equally to this work.

\section*{Competing interests}
The authors declare no competing interests.
		
\section*{Supplementary information}
The supplementary material for this article is available at \href{https://doi.org/10.1038/s41535-021-00419-1}{https://doi.org/10.1038/s41535-021-00419-1}.
	
\end{footnotesize}
\bibliographystyle{naturemag}
\bibliography{Torsello_npjQM_post_print.bib}
	
\end{document}